\documentclass[journal=nalefd,manuscript=letter]{achemso}

\usepackage[version=3]{mhchem} 

\author{Daniel B. Durham}
\affiliation{Materials Science Division, Argonne National Laboratory, Lemont, Illinois 60439, USA}%

\author{Manifa Noor}
\affiliation{Department of Materials Science and Engineering, University of Texas at Dallas, Richardson, Texas 75080, USA}

\author{Khandker Akif Aabrar}%
\affiliation{School of Electrical and Computer Engineering, Georgia Institute of Technology, Atlanta, Georgia 30332, USA}%

\author{Yuzi Liu}
\affiliation{Center for Nanoscale Materials, Argonne National Laboratory, Lemont, Illinois 60439, USA}%

\author{Suman Datta}%
\affiliation{School of Electrical and Computer Engineering, Georgia Institute of Technology, Atlanta, Georgia 30332, USA}%

\author{Kyeongjae Cho}
\affiliation{Department of Materials Science and Engineering, University of Texas at Dallas, Richardson, Texas 75080, USA}

\author{Supratik Guha}
\affiliation{Materials Science Division, Argonne National Laboratory, Lemont, Illinois 60439, USA}%
\alsoaffiliation{Pritzker School for Molecular Engineering, The University of Chicago, Chicago, Illinois 60637, USA}

\author{Charudatta Phatak}
\email{cd@anl.gov}
\affiliation{Materials Science Division, Argonne National Laboratory, Lemont, Illinois 60439, USA}%

\title[EPimaging]
  {Direct imaging of asymmetric interfaces and electrostatic potentials inside a hafnia-zirconia ferroelectric nanocapacitor}

\begin{document}

\begin{abstract}
  In hafnia-based thin-film ferroelectric devices, chemical phenomena during growth and processing such as oxygen vacancy formation and interfacial reactions appear to strongly affect device performance. However, the nanoscale structure, chemistry, and electrical potentials in these devices are not fully known, making it difficult to understand their influence on device properties. Here, we directly image the composition and electrostatic potential with nanometer resolution in the cross section of a nanocrystalline W / \ce{Hf_{0.5}Zr_{0.5}O_{2-\delta}} (HZO) / W ferroelectric capacitor using multimodal electron microscopy. This reveals a 1.4 nm wide tungsten sub-oxide interfacial layer formed at the bottom interface during fabrication which introduces a potential dip and leads to asymmetric switching fields. Additionally, the measured inner potential in HZO is consistent with the presence of about 20$\%$ oxygen vacancies and a negative built-in potential in HZO. These chemical and electrostatic details are important to characterize and tune to achieve high performance ferroelectric devices.
\end{abstract}

Ferroelectric transistors and capacitors based on hafnia and zirconia have become well established candidates for integration into the next generations of logic and memory.\cite{mueller2012incipient,salahuddin2018era,murray2018basic,haensch2022compute} The ferroelectricity of doped hafnia has been shown to be preserved down to single-nm film thicknesses.\cite{cheema2020enhanced} Implementation in ferroelectric-dielectric superlattices allows to achieve record-low effective oxide thickness through negative capacitance\cite{salahuddin2008use,cheema2022ultrathin} as well as high-endurance and linear analog memory.\cite{aabrar2021beol,aabrar2022beol} Additionally, the back-end-of-line processes already in use for hafnia and zirconia-based high-k gates can be readily adapted to make such ferroelectric devices.

However, one of the main challenges for these devices is stabilizing favorable chemical and electrostatic profiles across the thin-film stack. The performance of hafnia-based ferroelectric devices is highly sensitive to the choice of electrode materials and processing parameters. In many cases, devices experience high variability over cycling, such as requiring initial "wake-up" cycles to reach maximum polarization followed by gradual degradation and eventual failure.\cite{karbasian2017stabilization,cao2018effects,kashir2021large,wang2022enhanced,segantini2023interplay,alcala2023electrode,wang2024modulation} These effects are often attributed to charge trapping and diffusion of oxygen vacancies as well as chemical interdiffusion at interfaces. On the other hand, certain concentrations of oxygen vacancies are thought to help stabilize the ferroelectric orthorhombic phase, and in some cases, carefully tuning electrode oxidation appears to provide some performance gains, including enhanced remnant polarization, improved endurance, and reduced leakage.\cite{wang2024modulation,chiniwar2024ferroelectric} 

Tungsten electrodes have become a leading candidate thanks to demonstrated low leakage, low coercive field, and high endurance.\cite{cao2018effects,wang2022enhanced,wang2024modulation,chiniwar2024ferroelectric} However, achieving wake-up free devices with symmetric, low voltage switching and high endurance is still a challenge. It is still unclear to what extent oxygen vacancies in the ferroelectric benefit or hamper electrical response and what the nature and effects of the interfaces are. An important step to address this is to determine the composition and electrostatic potential profiles across the film stack and identify relationships to electrical properties.

Here, we directly image the chemical and electrostatic potential distributions across a W/\ce{Hf_{0.5}Zr_{0.5}O_{2-\delta}} (HZO)/W nanocapacitor using electron microscopy. In this case, we find \ce{O2} plasma-enhanced atomic layer deposition (PEALD) of the HZO introduces a thin \ce{WO_{x}} layer at the bottom interface, leading to a measurable potential dip as well as asymmetric electrical switching fields. By comparing the measured potentials with density functional theory calculations, we elucidate contributions from oxygen vacancies, the \ce{WO_{x}} layer, and a possible built-in electrical potential in HZO.

\begin{figure*}[ht]
    \centering
    \includegraphics[width=17.5cm]{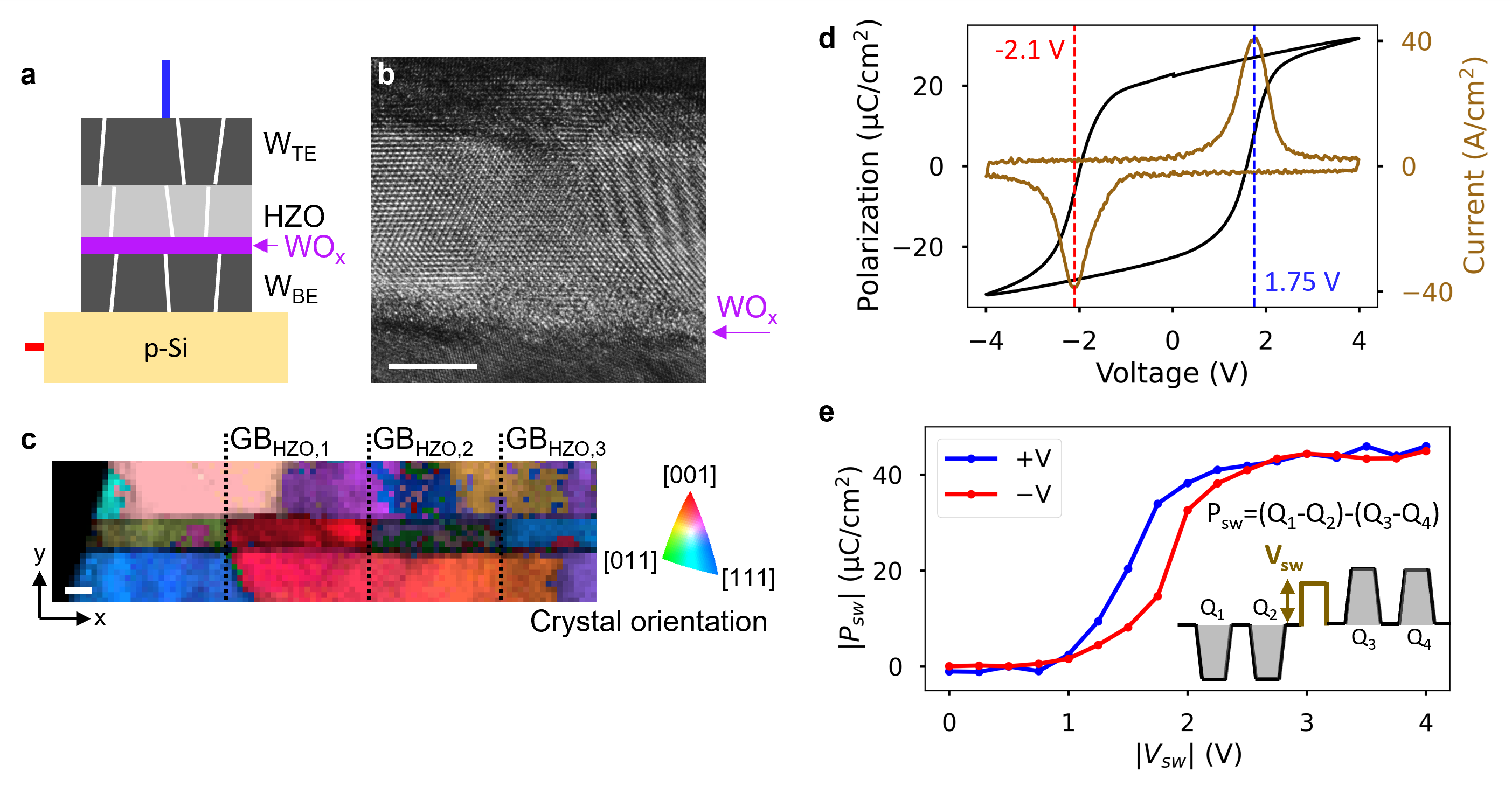}
    \caption{Asymmetric structure and switching of a nanocrystalline W/HZO/W ferroelectric capacitor. a) Schematic illustrating the device stack, including polycrystalline W top electrode ($\rm{W_{TE}}$) and bottom electrode ($\rm{W}_{BE}$), \ce{Hf_{0.5}Zr_{0.5}O_{2-\delta}} (HZO) ferroelectric layer, and tungsten oxide (\ce{WO_{x}}) interfacial layer formed during fabrication. 
    b) High-resolution TEM image of the cross section, showing polycrystalline HZO and the amorphous interfacial layer. Scale = 5 nm. c) Out-of-plane crystal orientation map obtained using scanning electron nanodiffraction. Scale = 10 nm. d) Polarization and current vs applied voltage during a 10 $\mathrm{\mu}$s rise time triangle wave voltage cycle. Asymmetric switching voltages are denoted by the vertical dashed lines. Positive polarity is defined by positive voltage at the top electrode. e) Switched polarization ($P_{sw}$) by 1 $\mathrm{\mu}$s wide square voltage pulses with amplitude $V_{sw}$ measured using the PUND method illustrated in the inset.}
    \label{fig:Sample}
\end{figure*}

The device examined in this study is detailed in Figure \ref{fig:Sample}. It is a W / \ce{Hf_{0.5}Zr_{0.5}O_{2-\delta}} / W (130 nm / 10 nm / 130 nm) ferroelectric capacitor deposited on p-doped Si. The HZO layer was deposited using \ce{O2} PEALD at 250$^\circ$C substrate temperature, while W bottom and top electrodes were deposited by sputtering. Afterwards, a rapid thermal anneal was performed at 400$^\circ$C in \ce{N2} environment for 60s to crystallize and stabilize the orthorhombic phase.\cite{aabrar2021beol} 

For TEM characterization, thin cross-section lamellae were lifted out using 30 kV focused ion beam milling and then polished with 500 V Ar ion milling to remove damaged surface layers. Devices were "woken up" before liftout by applying triangular voltage cycles with $\pm$4 V amplitude and 40 \textmu s period until a stable, single-loop hysteresis was observed (about 200 cycles). 

High-resolution TEM imaging (Figure \ref{fig:Sample}b) reveals that the HZO is polycrystalline with wide grains extending through most of its thickness. Scanning electron nanodiffraction with automated crystal orientation mapping\cite{ophus2022automated,savitzky2021py4dstem} (Figure \ref{fig:Sample}c, using a cubic approximation for HZO for simplicity) views this on a broader scale, showing grains up to 50 nm wide spanning most of the film thickness. However, in the high-resolution TEM, some structural disorder is visible at the bottom interface, suggesting the presence of a nanometer-scale interfacial layer.

The structural asymmetry leads to electrical switching asymmetry. Polarization and current loops from a woken-up device are shown in Figure \ref{fig:Sample}d. The switching voltage for positive polarity is 0.35 V lower than that for negative. Square pulse switching measurements in the PUND (positive-up-negative-down) scheme show a similar asymmetry in the threshold voltage. Such asymmetry can be problematic in thinner devices with sub-V switching: large shifts towards one polarity can increase the pulse amplitudes required for the same response and significantly reduce the remnant polarization.\cite{jiang2022enabling} So, it is important to understand and control the sources of this asymmetry.

\begin{figure}
    \centering
    \includegraphics[width=8.5cm]{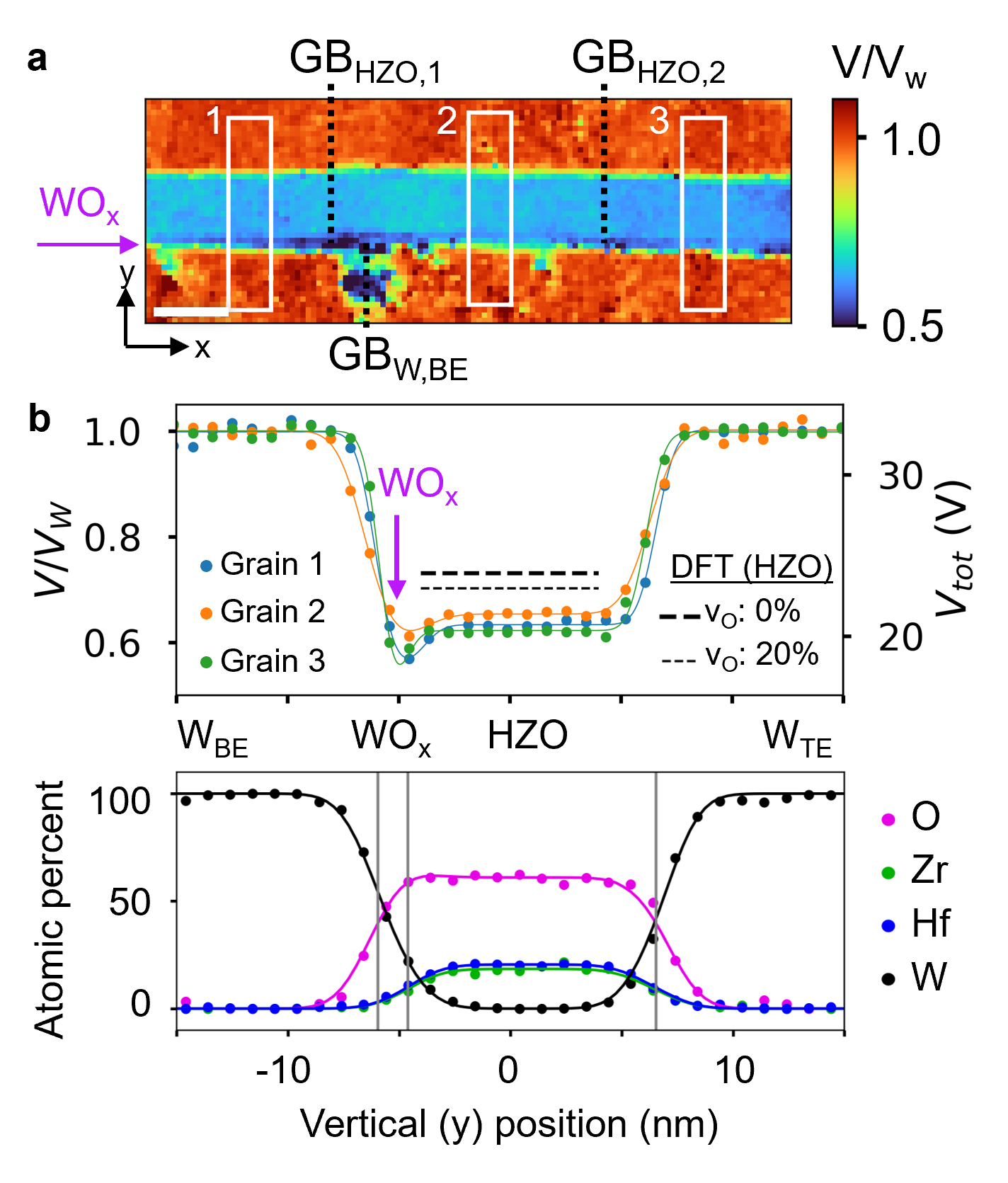}
    \caption{Electrostatic and chemical nanoimaging. a) Electrostatic potential image via off-axis electron holography, normalized with respect to the value in the W electrodes ($V/V_{W}$). The \ce{WO_{x}} layer at the bottom interface is labeled. $\rm{GB_{HZO}}$ and $\rm{GB_{W}}$ indicates lateral positions of HZO and W grain boundaries. The grain boundaries are the same as those indicated in Figure \ref{fig:Sample}c. Scale = 10 nm. b) (upper) Potential profiles across the film stack taken from three grains, i.e. the numbered regions in panel a. Right axis is the absolute potential assuming $V_{W} = 32.7$ V per our DFT calculation. Solid lines are a guide to the eye. Horizontal dashed lines (black) are the potentials in HZO predicted by our DFT calculations for W/HZO/W stacks wherein HZO contains 0$\%$ and 20$\%$ oxygen vacancies ($\rm{v_{O}}$). 
    (lower) Composition profile obtained using STEM EDS. Solid lines are fit profiles using a 4 layer model assuming the labeled components. The thin vertical lines indicate the fit layer boundaries.}
    \label{fig:Holo}
\end{figure}

To investigate the chemical and electrical features in this device, we performed electrostatic potential imaging using off-axis electron holography. Our experiments were done on an image-corrected JEOL 2100F TEM. By passing the electron beam at the sample edge and using an electrostatic biprism to interfere the part passing through the sample with that through vacuum, an interference pattern is generated from which the coherent phase shift, $\phi$, of the electron wave can be retrieved.\cite{joy1993practical} This $\phi$ can be related to the total electrostatic potential, $V_{tot}$, observed along the beam path (z direction): 
\begin{equation}
    \phi(x,y) = \sigma \int{V_{tot}(x,y,z) dz}
    \label{eq:phase}
\end{equation}
where $\sigma$ is the relativistic interaction parameter (7.29 mrad V$^{-1}$ nm$^{-1}$ for 200 keV electrons).\cite{Kirkland_2010} Importantly, this potential seen by the relativistic probe electrons differs from that experienced by charge carriers within the material. Here, $V_{tot}$ can be considered the sum of two components:
\begin{equation}
    V_{tot} = V^{0}+V_{e}
\end{equation}
$V^{0}$, the mean inner potential, largely originates from the core atomic potentials screened by diffuse electron clouds and reflects the material structure, electronic configuration, and composition. It is often tens of volts and strongly material dependent.\cite{kruse2006determination} $V_{e}$ is the potential from electric fields in the device such as from polarization, charge carriers, and built-in potentials.\cite{han2014interface,toyama2020quantitative,da2022assessment}

A representative electrostatic potential image normalized by the potential measured for W is shown in Figure \ref{fig:Holo}a. Here, the specimen was tilted by 4 degrees to minimize scattering artifacts from the middle grain, which was closely oriented to the [001] zone axis (see Figure \ref{fig:Sample}c). There is a strong contrast between $V_{tot}$ in the HZO and in W, mainly due to the V$^0$ difference. Additionally, a thin, low-potential layer is visible along the bottom interface. Selected line profiles from the three HZO grains in view (Figure \ref{fig:Holo}b) show that while there are slight variations in the measured HZO and interfacial potential, the dips at the bottom interface have similar depth across the grains. 

The chemical profile across the stack was examined using STEM energy-dispersive spectroscopy (EDS) (Figure \ref{fig:Holo}b). Signals were fit using a 4-layer film stack model to retrieve the widths and compositions. The retrieved composition of the HZO 
is 0.90:1 Zr:Hf and 1.56:1 O:(Hf,Zr), suggesting the presence of oxygen vacancies on the order of 20$\%$. This is consistent with XPS measurements of W-electrode HZO devices done by others which find $v_{O}$ concentrations in the range of 10-25\%.\cite{wang2024modulation} In addition, the O signal extends beyond the Hf and Zr signals into the bottom interface, implying a \ce{WO_{x}} interfacial layer. We find the \ce{WO_{x}} is 1.4 nm thick with x $\approx$ 2.5. This is likely formed during the PEALD of HZO, as the energetic \ce{O2} plasma can react with the bottom W surface to form the oxide.\cite{chiniwar2024ferroelectric}

To determine the origins of the observed potential profile, we performed density functional theory calculations of the electrostatic potentials in ideal epitaxial W(110)/HZO(001)/W(110) stacks as well as individual material slabs for W, HZO, and \ce{WO_{x}}. This provides mainly the inner potentials, $V^{0}$, though certain electronic contributions such as polarization also appear to be present as will be shown. The average potentials for the material combinations examined here are listed in Table~\ref{tab:pots}.

\begin{table}[h!]
    \centering
    \begin{tabular}{l|c|c}
         Configuration & $V^{0}$ (V) & $V/V_{W}$ \\
         \hline
         W & 32.7 & 1 \\
         m-HZO & 23.7 & 0.725 \\
         o-HZO & 24.0 & 0.734 \\
         o-HZO sandwich & 23.9 & 0.731 \\
         o-HZO sandwich (20\% $\rm{v_{O}}$) & 23.0 & 0.703 \\
         \ce{t-WO2} & 20.6 & 0.630 \\
         \ce{t-WO3} & 19.6 & 0.599 \\
    \end{tabular}
    \caption{Mean inner potentials calculated using DFT. Both the absolute inner potential with respect to vacuum ($V^{0}$) and the ratio with respect to W ($V/V_{W}$) are reported. o- is orthorhombic, m- is monoclinic, t- is tetragonal. Sandwich indicates W/HZO/W heterostructure. $\rm{v_O}$ indicates oxygen vacancies, which are only included where explicitly noted.}
    \label{tab:pots}
\end{table}

\begin{figure}[ht]
    \centering
    \includegraphics[width=8.5cm]{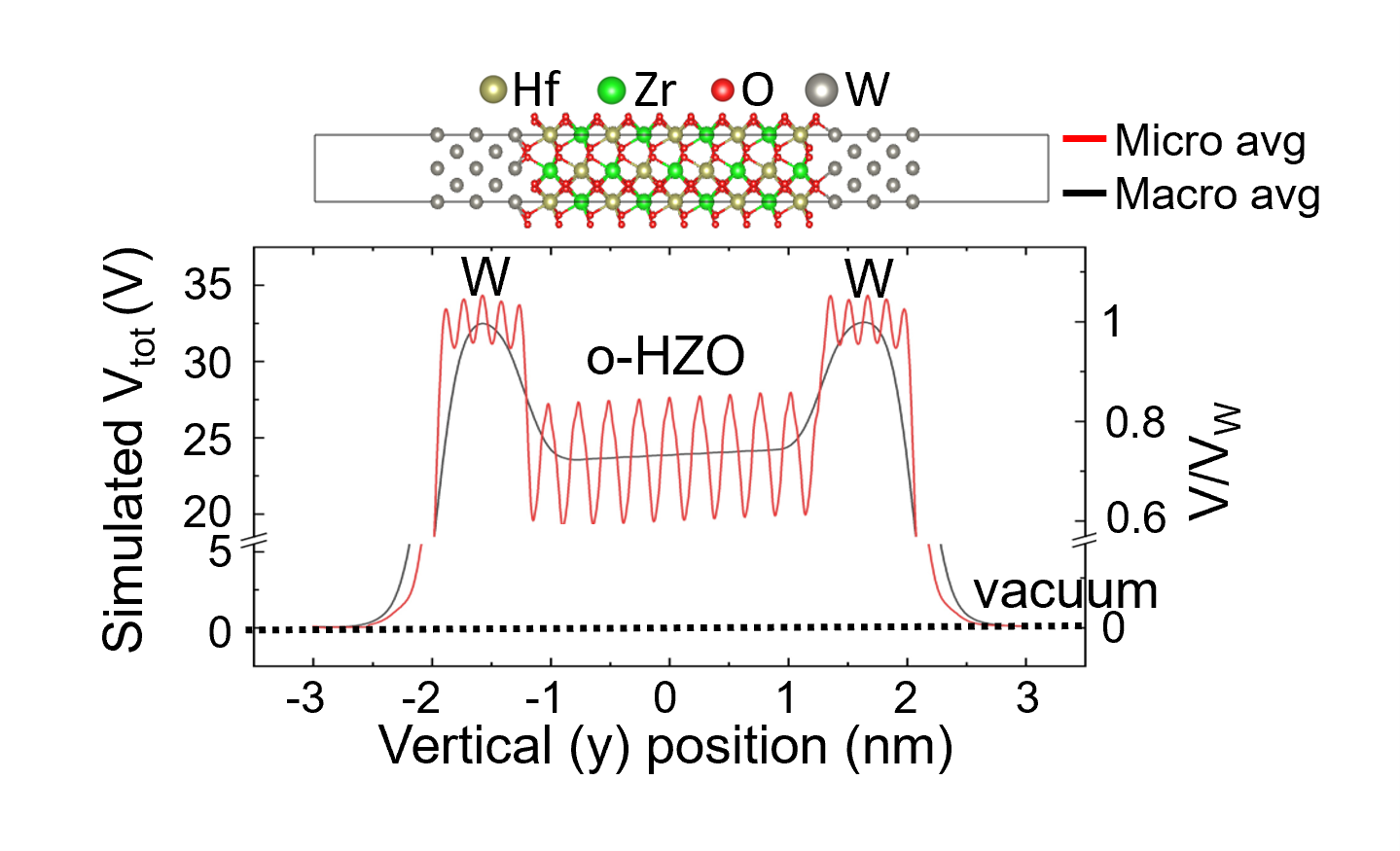}
    \caption{Density functional theory (DFT) calculation of the total electrostatic potential ($V_{tot}$)  profile across an idealized, fully polarized W/o-HZO/W stack (o- is the polar orthorhombic phase). Microscopic (micro) average is within layers only. Macroscopic average also averages across each atomic layer.}
    \label{fig:DFT}
\end{figure}

We first examine the predicted inner potential profile for a fully polarized, orthorhombic \ce{Hf_{0.5}Zr_{0.5}O_{2}} layer between W electrodes, shown in Figure \ref{fig:DFT}. The W and HZO layers show large, distinct average potentials with respect to vacuum. In the stack, $V^{0}_{W} =$ 32.7 V, while $V^{0}_{HZO} =$ 24.0 V. Comparing to individual material slabs for W and HZO, we find the effect of interfacing the materials in these simulations is negligible, less than 1$\%$. This suggests that interface dipoles in the simulated structure are negligible. Polarization also has little effect on the mean value, as the polar o-HZO and non-polar m-HZO phases also differ by less than 1$\%$. That said, a fully polarized W/HZO/W stack does show a large internal field of 0.38 V nm$^{-1}$, suggesting polarization could in principle also be detected via electrostatic potential imaging. However, in order to reliably detect such a gradient from experimental images, we would need to perform in situ experiments where the polarization is switched to distinguish it from other effects such as thickness gradients. In real nanocrystalline devices, this may be difficult to observe as the grains are randomly oriented, not necessarily fully switched, and these fields may be screened by additional free and trapped charge carriers.

The DFT-calculated inner potentials can then be compared with the measured potentials to determine the chemical and electronic contributions. The measured $V_{HZO}/V_{W}$ is 0.641 $\pm$ 0.016, which is significantly lower than the 0.731 value predicted by DFT for the pure W/o-HZO/W stack. Assuming $V_{W}$ is accurate, then this implies $V_{HZO} = 21.0$ V, which is 2.9 V lower than the 23.9 V predicted for pure HZO. In part, this can be explained by reduction of the inner potential due to oxygen vacancies. We found that introducing 20 at\% neutral $v_{O}$ into the DFT model reduces the inner potential of the HZO by about 0.9 V. 

However, a 2.0 V reduction in experimentally measured HZO potential still remains, which could be due to a negative electrical potential ($V_{e}$). A possible explanation is formation of built-in potential in the HZO through band bending. This can occur when the ferroelectric has a high defect concentration: For instance, PZT has been found to show p-type character due to its vacancy defect chemistry.\cite{pintilie2005metal,han2014interface} Considering that the bandgaps of hafnia and zirconia films are typically around 5.8 to 6.0 V and that the work function of W aligns closely with the middle of these gaps,\cite{nguyen2005sub} then a -2.0 V built-in potential is feasible if the HZO  is effectively n-type. Oxygen vacancies could easily supply sufficient defect levels (20 at\% $v_{O}$ = 1.16 $\times$ 10$^{-20}$ cm$^{-3}$) and there is both theoretical and experimental evidence that oxygen vacancies introduce shallow levels in crystalline hafnia-zirconia materials and trap electrons therein.\cite{nguyen2005sub,ramo2007spectroscopic,islamov2019identification}

The potential dip at the bottom interface appears to be well explained by the inner potential of a \ce{WO_{x}} layer. For instance, using the tetragonal crystal structures for \ce{WO2} and \ce{WO3}, we find $V/V_{W}$ to be 0.630 and 0.599 respectively, which is close to the experimental values that range from 0.55 to 0.60. This difference may be due to the amorphous nature of the observed film compared to the crystalline models used here.

Altogether, the imaging experiments and DFT calculations reported here illuminate the complex chemical and electrostatic profiles in a hafnia-zirconia ferroelectric capacitor. Though W electrodes are generally regarded as less chemically reactive than some common alternatives such as TiN, we find that nm-scale \ce{WO_{x}} interfaces can be formed, for instance using \ce{O2} plasma-enhanced ALD, which impart asymmetry in the electrical response. In addition, we find evidence of $\approx$ 20 \% oxygen vacancies in the HZO and a negative built-in potential which suggests the O-deficient HZO shows effectively n-type character. This affects how to engineer material and interface work functions to tune barrier heights to reduce leakage and charge trapping. These potential measurements and simulations provide a foundation for future investigations of how the potential profiles depend on ferroelectric, intefacial layer, and electrode compositions, as well as how the potentials change during electrical cycling. Understanding and controlling the materials and interfaces across the entire film stack is necessary to minimize leakage, maximize endurance, and achieve zero wake-up in next-generation ferroelectric devices.

\begin{acknowledgement}

This material is based upon work supported by the U.S. Department of Energy, Office of Science, for support of microelectronics research, under contract number DE-AC0206CH11357. Work performed at the Center for Nanoscale Materials, a U.S. Department of Energy Office of Science User Facility, was supported by the U.S. DOE, Office of Basic Energy Sciences, under Contract No. DE-AC02-06CH11357. M.N. and K.C. were supported by Semiconductor Research Corporation (SRC), grant number 3146.001. S.D. and K. A.A. from Georgia Tech acknowledge support for the growth of ferroelectric films from Department of Energy, Atoms-to-Systems Co-Design: Transforming Data Flow to Accelerate Scientific Discovery project at SLAC National Accelerator Laboratory.

The submitted manuscript has been created by UChicago Argonne, LLC, Operator of Argonne National Laboratory (“Argonne”). Argonne, a U.S. Department of Energy Office of Science laboratory, is operated under Contract No. DE-AC02-06CH11357. The U.S. Government retains for itself, and others acting on its behalf, a paid-up nonexclusive, irrevocable worldwide license in said article to reproduce, prepare derivative works, distribute copies to the public, and perform publicly and display publicly, by or on behalf of the Government. The Department of Energy will provide public access to these results of federally sponsored research in accordance with the DOE Public Access Plan. http://energy.gov/downloads/doe-public-access-plan

\end{acknowledgement}

\renewcommand{\thefigure}{S\arabic{figure}}
\setcounter{figure}{0}

\newpage
\section{Methods}

\subsection{PUND measurements}

Positive-up-negative-down (PUND) measurements were performed by measuring the integrated charge density over two sequential pulses with opposing polarity to the switching pulse (Q1, Q2), followed by measuring the integrated over two sequential pulses with matching polarity to the switching pulse (Q3, Q4) and then retrieving the switched polarization as (Q1-Q2)-(Q3-Q4). This is illustrated by the inset in Figure 1 of the main text. The Q1-Q4 pulses had 4V amplitude with 10 $\mu$s duration, rise time, and fall time. Switching pulses had 1 $\mu$s duration and 50 ns rise and fall times. There was 20 $\mu$s spacing between the end of each pulse and the start of the next.

\subsection{DFT electrostatic potential calculations}

Vienna ab initio software package (VASP)\cite{kresse1996efficiency,kresse1996efficient} was used with projected augmented wave (PAW)\cite{blochl1994projector,kresse1999ultrasoft} pseudopotentials. A vacuum length of 10 \AA  ~was included on both sides to create a slab geometry. Exchange and correlation interactions were included by employing the Perdew–Burke–Ernzerhof (PBE)\cite{perdew1996generalized} functional. The kinetic energy cutoff of 520 eV was used for plane wave basis set. Brillouin zone sampling was performed by Monkhorst-Pack\cite{monkhorst1976special} k-point mesh with sampling density of 0.25 2$\pi$\AA$^{-1}$. The converged energy criterion for self-consistent field minimization of electronic structure was 10$^{-7}$ eV. The macroscopic and planar average electrostatic potential were calculated using VASPKIT.\cite{wang2021vaspkit} 

\subsection{EDS fitting}

Composition was extracted by fitting the EDS signal profiles with a model assuming 4 rectangular layers (W, \ce{WO_{x}}, \ce{HfZr_{x}O_{y}}, and W) convolved with a gaussian beam size. This fitting is illustrated in Supporting Figure~\ref{fig:EDSFit}. The compositions of each layer in atomic percent were then computed in Hyperspy\cite{hyperspy} using the absorption-corrected Cliff-Lorimer method (using the k factors provided in Bruker Espirit). For the absorption correction, we used the average lamella thickness in the measured region of 53 nm, which was determined using the phase signal in W assuming $V_{W}$ = 32.7 V as predicted by our DFT calculations.

\newpage

\section{Supporting Figures}

\begin{figure}[h!]
    \centering
    \includegraphics[width=11.0cm]{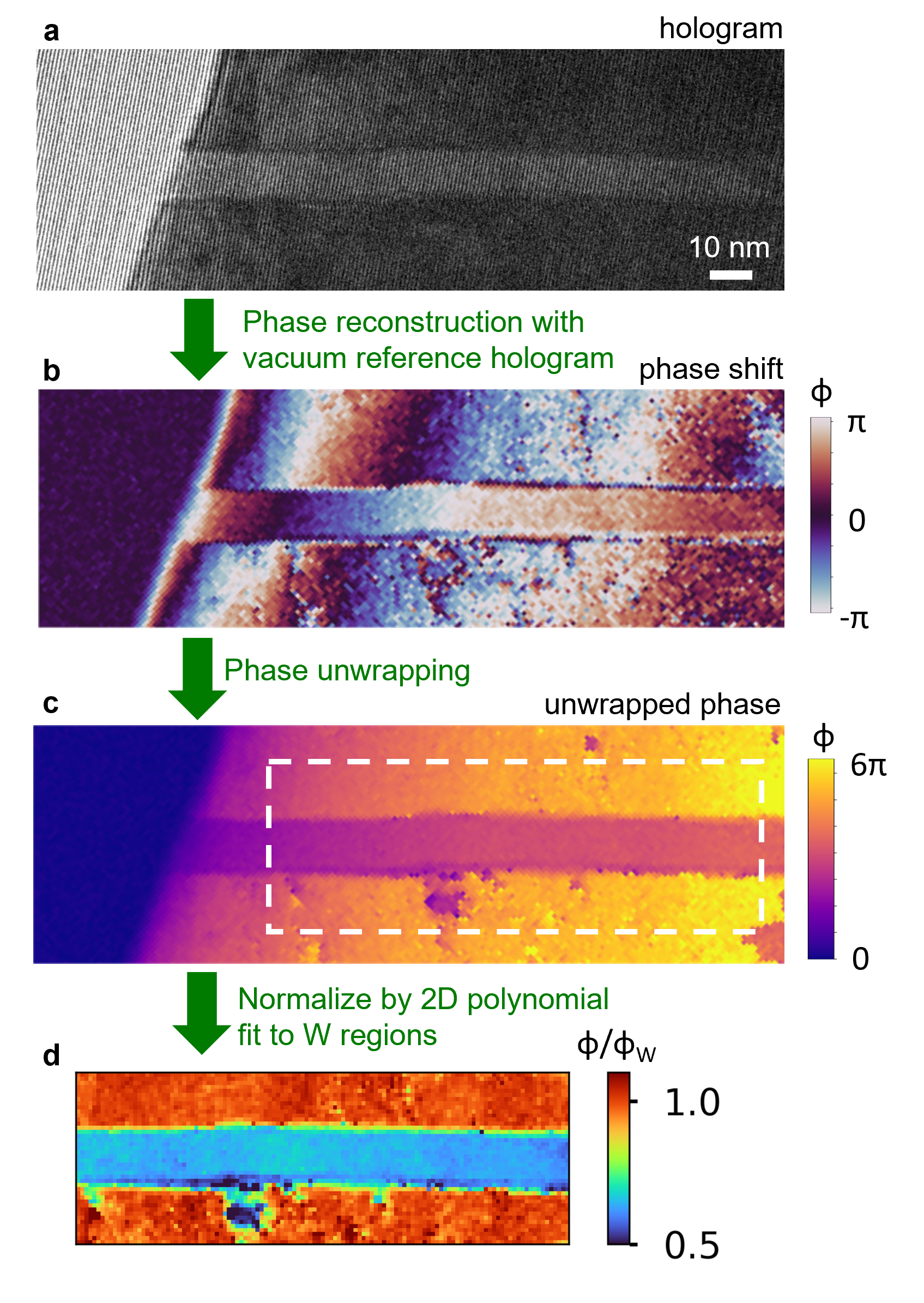}
    \caption{Phase and potential reconstruction via off-axis electron holography. a) Fringe pattern. b) Reconstructed phase after processing with reference hologram in vacuum. c) Unwrapped phase image. d) 2D polynomial fit to W regions}
    \label{fig:HoloProcess}
\end{figure}

\newpage

\begin{figure}[h!]
    \centering
    \includegraphics[width=12.0cm]{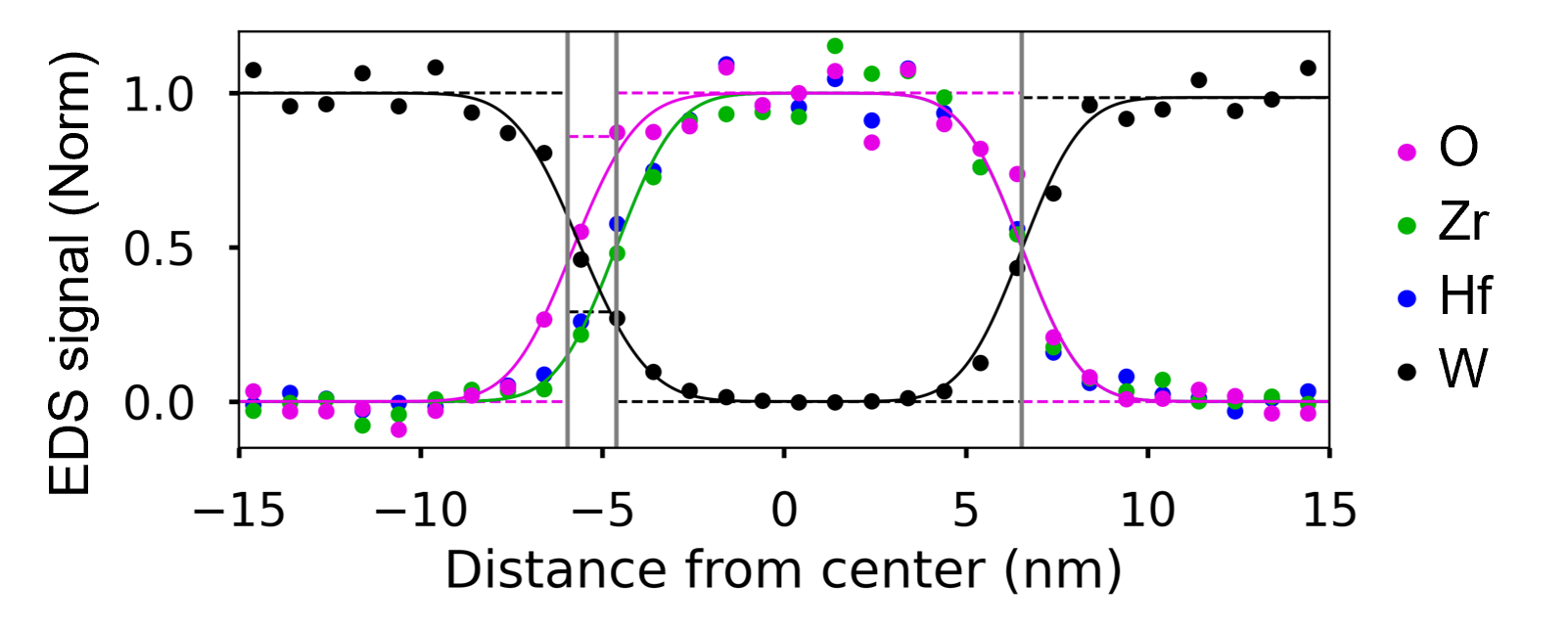}
    \caption{Fitting of normalized EDS signals with 4-layer model convolved with electron beam size. Solid lines are the fit profiles. Vertical lines indicate fit layer edges. Dashed horizontal lines indicate EDS signal levels in each layer. These levels were used to compute the atomic percent of each layer.}
    \label{fig:EDSFit}
\end{figure}

\newpage

\begin{figure}[h!]
    \centering
    \includegraphics[width=15.0cm]{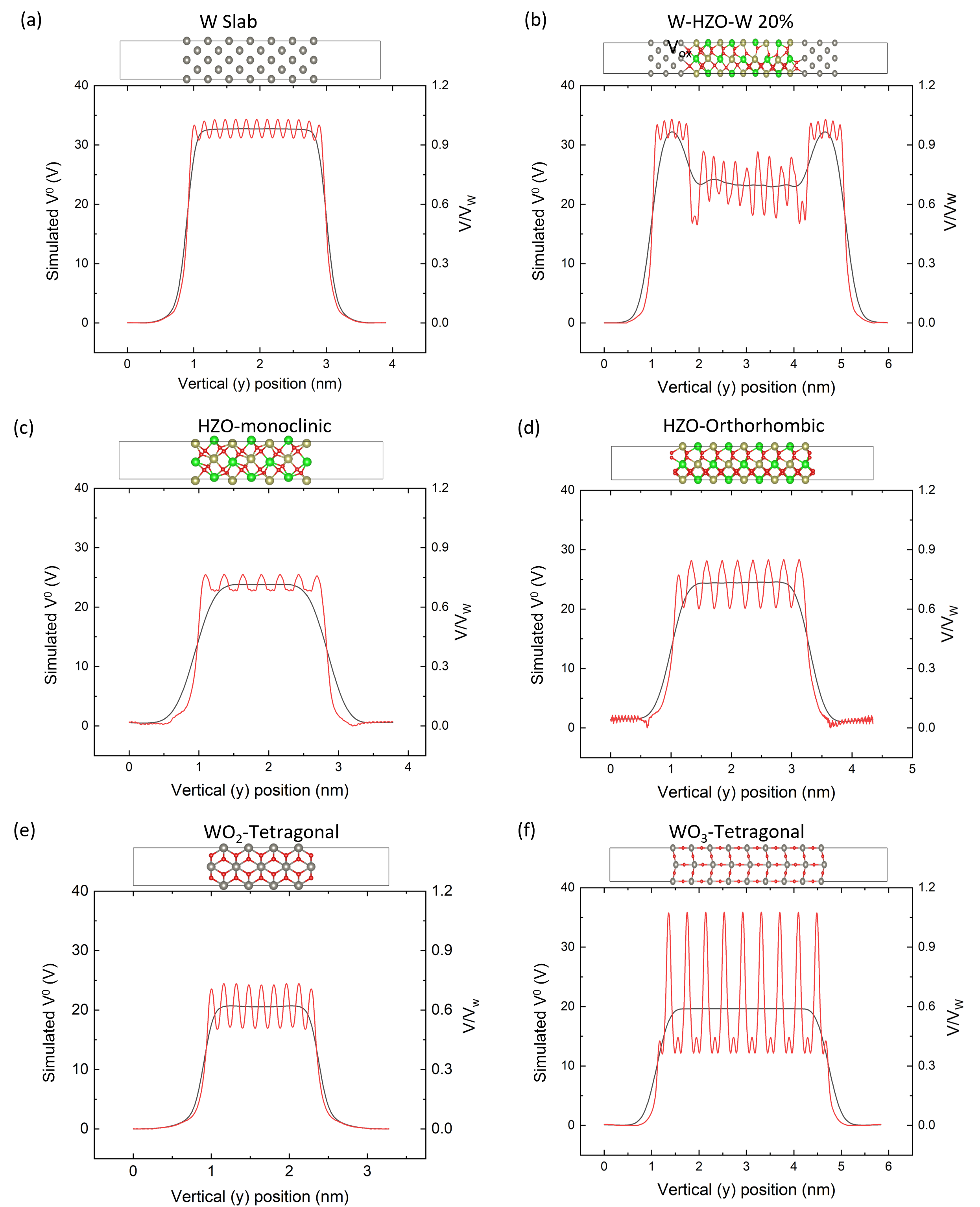}
    \caption{DFT calculations of the electrostatic potential ($V^{0}$) profile of (a) W (110) metal, (b) W/o-HZO/W with 20\% oxygen vacancy, (c) m-HZO, (d) o-HZO, (e) t-WO2 and (f) t-WO3 slab models. Left Here o-, m- and t- indicate orthorhombic, monoclinic, and tetragonal phases respectively. Atom colors correspond to W (gray), Hf (yellow), Zr (green), and O (red). Microscopic (micro) average (red lines) is within layers only. Macroscopic average (black lines) also averages across each atomic layer.}
    \label{fig:DFTPlots}
\end{figure}

\newpage


\newpage

\bibliography{main}

\providecommand{\latin}[1]{#1}
\makeatletter
\providecommand{\doi}
  {\begingroup\let\do\@makeother\dospecials
  \catcode`\{=1 \catcode`\}=2 \doi@aux}
\providecommand{\doi@aux}[1]{\endgroup\texttt{#1}}
\makeatother
\providecommand*\mcitethebibliography{\thebibliography}
\csname @ifundefined\endcsname{endmcitethebibliography}  {\let\endmcitethebibliography\endthebibliography}{}
\begin{mcitethebibliography}{39}
\providecommand*\natexlab[1]{#1}
\providecommand*\mciteSetBstSublistMode[1]{}
\providecommand*\mciteSetBstMaxWidthForm[2]{}
\providecommand*\mciteBstWouldAddEndPuncttrue
  {\def\EndOfBibitem{\unskip.}}
\providecommand*\mciteBstWouldAddEndPunctfalse
  {\let\EndOfBibitem\relax}
\providecommand*\mciteSetBstMidEndSepPunct[3]{}
\providecommand*\mciteSetBstSublistLabelBeginEnd[3]{}
\providecommand*\EndOfBibitem{}
\mciteSetBstSublistMode{f}
\mciteSetBstMaxWidthForm{subitem}{(\alph{mcitesubitemcount})}
\mciteSetBstSublistLabelBeginEnd
  {\mcitemaxwidthsubitemform\space}
  {\relax}
  {\relax}

\bibitem[Mueller \latin{et~al.}(2012)Mueller, Mueller, Singh, Riedel, Sundqvist, Schroeder, and Mikolajick]{mueller2012incipient}
Mueller,~S.; Mueller,~J.; Singh,~A.; Riedel,~S.; Sundqvist,~J.; Schroeder,~U.; Mikolajick,~T. Incipient ferroelectricity in Al-doped \ce{HfO2} thin films. \emph{Advanced Functional Materials} \textbf{2012}, \emph{22}, 2412--2417\relax
\mciteBstWouldAddEndPuncttrue
\mciteSetBstMidEndSepPunct{\mcitedefaultmidpunct}
{\mcitedefaultendpunct}{\mcitedefaultseppunct}\relax
\EndOfBibitem
\bibitem[Salahuddin \latin{et~al.}(2018)Salahuddin, Ni, and Datta]{salahuddin2018era}
Salahuddin,~S.; Ni,~K.; Datta,~S. The era of hyper-scaling in electronics. \emph{Nature Electronics} \textbf{2018}, \emph{1}, 442--450\relax
\mciteBstWouldAddEndPuncttrue
\mciteSetBstMidEndSepPunct{\mcitedefaultmidpunct}
{\mcitedefaultendpunct}{\mcitedefaultseppunct}\relax
\EndOfBibitem
\bibitem[Murray \latin{et~al.}()Murray, Guha, Reed, Herrera, Kleese~van Dam, Salahuddin, Ang, Conte, Jena, Kaplar, \latin{et~al.} others]{murray2018basic}
Murray,~C.; Guha,~S.; Reed,~D.; Herrera,~G.; Kleese~van Dam,~K.; Salahuddin,~S.; Ang,~J.; Conte,~T.; Jena,~D.; Kaplar,~R.; others \emph{{Basic Research Needs for Microelectronics: Report of the Office of Science Workshop on Basic Research Needs for Microelectronics, October 23--25, 2018}}\relax
\mciteBstWouldAddEndPuncttrue
\mciteSetBstMidEndSepPunct{\mcitedefaultmidpunct}
{\mcitedefaultendpunct}{\mcitedefaultseppunct}\relax
\EndOfBibitem
\bibitem[Haensch \latin{et~al.}(2022)Haensch, Raghunathan, Roy, Chakrabarti, Phatak, Wang, and Guha]{haensch2022compute}
Haensch,~W.; Raghunathan,~A.; Roy,~K.; Chakrabarti,~B.; Phatak,~C.~M.; Wang,~C.; Guha,~S. Compute in-Memory with Non-Volatile Elements for Neural Networks: A Review from a Co-Design Perspective. \emph{Advanced Materials} \textbf{2022}, 2204944\relax
\mciteBstWouldAddEndPuncttrue
\mciteSetBstMidEndSepPunct{\mcitedefaultmidpunct}
{\mcitedefaultendpunct}{\mcitedefaultseppunct}\relax
\EndOfBibitem
\bibitem[Cheema \latin{et~al.}(2020)Cheema, Kwon, Shanker, Dos~Reis, Hsu, Xiao, Zhang, Wagner, Datar, McCarter, \latin{et~al.} others]{cheema2020enhanced}
Cheema,~S.~S.; Kwon,~D.; Shanker,~N.; Dos~Reis,~R.; Hsu,~S.-L.; Xiao,~J.; Zhang,~H.; Wagner,~R.; Datar,~A.; McCarter,~M.~R.; others Enhanced ferroelectricity in ultrathin films grown directly on silicon. \emph{Nature} \textbf{2020}, \emph{580}, 478--482\relax
\mciteBstWouldAddEndPuncttrue
\mciteSetBstMidEndSepPunct{\mcitedefaultmidpunct}
{\mcitedefaultendpunct}{\mcitedefaultseppunct}\relax
\EndOfBibitem
\bibitem[Salahuddin and Datta(2008)Salahuddin, and Datta]{salahuddin2008use}
Salahuddin,~S.; Datta,~S. Use of negative capacitance to provide voltage amplification for low power nanoscale devices. \emph{Nano letters} \textbf{2008}, \emph{8}, 405--410\relax
\mciteBstWouldAddEndPuncttrue
\mciteSetBstMidEndSepPunct{\mcitedefaultmidpunct}
{\mcitedefaultendpunct}{\mcitedefaultseppunct}\relax
\EndOfBibitem
\bibitem[Cheema \latin{et~al.}(2022)Cheema, Shanker, Wang, Hsu, Hsu, Liao, San~Jose, Gomez, Chakraborty, Li, \latin{et~al.} others]{cheema2022ultrathin}
Cheema,~S.~S.; Shanker,~N.; Wang,~L.-C.; Hsu,~C.-H.; Hsu,~S.-L.; Liao,~Y.-H.; San~Jose,~M.; Gomez,~J.; Chakraborty,~W.; Li,~W.; others Ultrathin ferroic \ce{HfO2}--\ce{ZrO2} superlattice gate stack for advanced transistors. \emph{Nature} \textbf{2022}, \emph{604}, 65--71\relax
\mciteBstWouldAddEndPuncttrue
\mciteSetBstMidEndSepPunct{\mcitedefaultmidpunct}
{\mcitedefaultendpunct}{\mcitedefaultseppunct}\relax
\EndOfBibitem
\bibitem[Aabrar \latin{et~al.}(2021)Aabrar, Gomez, Kirtania, San~Jose, Luo, Ravikumar, Ravindran, Ye, Banerjee, Dutta, Khan, Yu, and Datta]{aabrar2021beol}
Aabrar,~K.~A.; Gomez,~J.; Kirtania,~S.~G.; San~Jose,~M.; Luo,~Y.; Ravikumar,~P.~G.; Ravindran,~P.~V.; Ye,~H.; Banerjee,~S.; Dutta,~S.; Khan,~A.~I.; Yu,~S.; Datta,~S. {BEOL} compatible superlattice {FerroFET}-based high precision analog weight cell with superior linearity and symmetry. 2021 IEEE International Electron Devices Meeting (IEDM). 2021; pp 19--6\relax
\mciteBstWouldAddEndPuncttrue
\mciteSetBstMidEndSepPunct{\mcitedefaultmidpunct}
{\mcitedefaultendpunct}{\mcitedefaultseppunct}\relax
\EndOfBibitem
\bibitem[Aabrar \latin{et~al.}(2022)Aabrar, Kirtania, Liang, Gomez, San~Jose, Luo, Ye, Dutta, Ravikumar, Ravindran, Khan, Yu, and Datta]{aabrar2022beol}
Aabrar,~K.~A.; Kirtania,~S.~G.; Liang,~F.-X.; Gomez,~J.; San~Jose,~M.; Luo,~Y.; Ye,~H.; Dutta,~S.; Ravikumar,~P.~G.; Ravindran,~P.~V.; Khan,~A.~I.; Yu,~S.; Datta,~S. {BEOL}-compatible superlattice {FEFET} analog synapse with improved linearity and symmetry of weight update. \emph{IEEE Transactions on Electron Devices} \textbf{2022}, \emph{69}, 2094--2100\relax
\mciteBstWouldAddEndPuncttrue
\mciteSetBstMidEndSepPunct{\mcitedefaultmidpunct}
{\mcitedefaultendpunct}{\mcitedefaultseppunct}\relax
\EndOfBibitem
\bibitem[Karbasian \latin{et~al.}(2017)Karbasian, Dos~Reis, Yadav, Tan, Hu, and Salahuddin]{karbasian2017stabilization}
Karbasian,~G.; Dos~Reis,~R.; Yadav,~A.~K.; Tan,~A.~J.; Hu,~C.; Salahuddin,~S. Stabilization of ferroelectric phase in tungsten capped \ce{Hf_{0.8}Zr_{0.2}O_{2}}. \emph{Applied Physics Letters} \textbf{2017}, \emph{111}\relax
\mciteBstWouldAddEndPuncttrue
\mciteSetBstMidEndSepPunct{\mcitedefaultmidpunct}
{\mcitedefaultendpunct}{\mcitedefaultseppunct}\relax
\EndOfBibitem
\bibitem[Cao \latin{et~al.}(2018)Cao, Wang, Zhao, Yang, Zhao, Wang, Zhang, Lv, Liu, and Liu]{cao2018effects}
Cao,~R.; Wang,~Y.; Zhao,~S.; Yang,~Y.; Zhao,~X.; Wang,~W.; Zhang,~X.; Lv,~H.; Liu,~Q.; Liu,~M. Effects of capping electrode on ferroelectric properties of \ce{Hf_{0.5} Zr_{0.5} O_{2}} thin films. \emph{IEEE Electron Device Letters} \textbf{2018}, \emph{39}, 1207--1210\relax
\mciteBstWouldAddEndPuncttrue
\mciteSetBstMidEndSepPunct{\mcitedefaultmidpunct}
{\mcitedefaultendpunct}{\mcitedefaultseppunct}\relax
\EndOfBibitem
\bibitem[Kashir \latin{et~al.}(2021)Kashir, Kim, Oh, and Hwang]{kashir2021large}
Kashir,~A.; Kim,~H.; Oh,~S.; Hwang,~H. Large remnant polarization in a wake-up free \ce{Hf_{0.5}Zr_{0.5}O_{2}} ferroelectric film through bulk and interface engineering. \emph{ACS Applied Electronic Materials} \textbf{2021}, \emph{3}, 629--638\relax
\mciteBstWouldAddEndPuncttrue
\mciteSetBstMidEndSepPunct{\mcitedefaultmidpunct}
{\mcitedefaultendpunct}{\mcitedefaultseppunct}\relax
\EndOfBibitem
\bibitem[Wang \latin{et~al.}(2022)Wang, Zhang, Wang, Luo, Li, Shuai, Tao, Fan, Chen, Zeng, Dai, Lu, and Liu]{wang2022enhanced}
Wang,~D.; Zhang,~Y.; Wang,~J.; Luo,~C.; Li,~M.; Shuai,~W.; Tao,~R.; Fan,~Z.; Chen,~D.; Zeng,~M.; Dai,~J.~Y.; Lu,~X.~B.; Liu,~J.-M. Enhanced ferroelectric polarization with less wake-up effect and improved endurance of \ce{Hf_{0.5}Zr_{0.5}O_{2}} thin films by implementing W electrode. \emph{Journal of Materials Science \& Technology} \textbf{2022}, \emph{104}, 1--7\relax
\mciteBstWouldAddEndPuncttrue
\mciteSetBstMidEndSepPunct{\mcitedefaultmidpunct}
{\mcitedefaultendpunct}{\mcitedefaultseppunct}\relax
\EndOfBibitem
\bibitem[Segantini \latin{et~al.}(2023)Segantini, Manchon, Ca{\~n}ero~Infante, Bugnet, Barhoumi, Nirantar, Mayes, Rojo~Romeo, Blanchard, Deleruyelle, Sriram, and Vilquin]{segantini2023interplay}
Segantini,~G.; Manchon,~B.; Ca{\~n}ero~Infante,~I.; Bugnet,~M.; Barhoumi,~R.; Nirantar,~S.; Mayes,~E.; Rojo~Romeo,~P.; Blanchard,~N.; Deleruyelle,~D.; Sriram,~S.; Vilquin,~B. Interplay between Strain and Defects at the Interfaces of Ultra-Thin \ce{Hf_{0.5}Zr_{0.5}O_{2}}-Based Ferroelectric Capacitors. \emph{Advanced Electronic Materials} \textbf{2023}, \emph{9}, 2300171\relax
\mciteBstWouldAddEndPuncttrue
\mciteSetBstMidEndSepPunct{\mcitedefaultmidpunct}
{\mcitedefaultendpunct}{\mcitedefaultseppunct}\relax
\EndOfBibitem
\bibitem[Alcala \latin{et~al.}(2023)Alcala, Materano, Lomenzo, Vishnumurthy, Hamouda, Dubourdieu, Kersch, Barrett, Mikolajick, and Schroeder]{alcala2023electrode}
Alcala,~R.; Materano,~M.; Lomenzo,~P.~D.; Vishnumurthy,~P.; Hamouda,~W.; Dubourdieu,~C.; Kersch,~A.; Barrett,~N.; Mikolajick,~T.; Schroeder,~U. The Electrode-Ferroelectric Interface as the Primary Constraint on Endurance and Retention in HZO-Based Ferroelectric Capacitors. \emph{Advanced Functional Materials} \textbf{2023}, \emph{33}, 2303261\relax
\mciteBstWouldAddEndPuncttrue
\mciteSetBstMidEndSepPunct{\mcitedefaultmidpunct}
{\mcitedefaultendpunct}{\mcitedefaultseppunct}\relax
\EndOfBibitem
\bibitem[Wang \latin{et~al.}(2024)Wang, Slesazeck, Mikolajick, and Grube]{wang2024modulation}
Wang,~X.; Slesazeck,~S.; Mikolajick,~T.; Grube,~M. Modulation of Oxygen Content and Ferroelectricity in Sputtered Hafnia-Zirconia by Engineering of Tungsten Oxide Bottom Electrodes. \emph{Advanced Electronic Materials} \textbf{2024}, 2300798\relax
\mciteBstWouldAddEndPuncttrue
\mciteSetBstMidEndSepPunct{\mcitedefaultmidpunct}
{\mcitedefaultendpunct}{\mcitedefaultseppunct}\relax
\EndOfBibitem
\bibitem[Chiniwar \latin{et~al.}(2024)Chiniwar, Hsieh, Shih, Teng, Yang, Hu, Lin, Tang, and Tseng]{chiniwar2024ferroelectric}
Chiniwar,~S.~P.; Hsieh,~Y.-C.; Shih,~C.-H.; Teng,~C.-Y.; Yang,~J.-L.; Hu,~C.; Lin,~B.-H.; Tang,~M.-T.; Tseng,~Y.-C. Ferroelectric Enhancement in a TiN/\ce{Hf_{1--x} Zr_{x} O_{2}}/W Device with Controlled Oxidation of the Bottom Electrode. \emph{ACS Applied Electronic Materials} \textbf{2024}, \relax
\mciteBstWouldAddEndPunctfalse
\mciteSetBstMidEndSepPunct{\mcitedefaultmidpunct}
{}{\mcitedefaultseppunct}\relax
\EndOfBibitem
\bibitem[Ophus \latin{et~al.}(2022)Ophus, Zeltmann, Bruefach, Rakowski, Savitzky, Minor, and Scott]{ophus2022automated}
Ophus,~C.; Zeltmann,~S.~E.; Bruefach,~A.; Rakowski,~A.; Savitzky,~B.~H.; Minor,~A.~M.; Scott,~M.~C. Automated crystal orientation mapping in {py4DSTEM} using sparse correlation matching. \emph{Microscopy and Microanalysis} \textbf{2022}, \emph{28}, 390--403\relax
\mciteBstWouldAddEndPuncttrue
\mciteSetBstMidEndSepPunct{\mcitedefaultmidpunct}
{\mcitedefaultendpunct}{\mcitedefaultseppunct}\relax
\EndOfBibitem
\bibitem[Savitzky \latin{et~al.}(2021)Savitzky, Zeltmann, Hughes, Brown, Zhao, Pelz, Pekin, Barnard, Donohue, DaCosta, Kennedy, Xie, Janish, Schneider, Herring, Gopal, Anapolsky, Dhall, Bustillo, Ercius, Scott, Ciston, Minor, and Ophus]{savitzky2021py4dstem}
Savitzky,~B.~H. \latin{et~al.}  {py4DSTEM}: A software package for four-dimensional scanning transmission electron microscopy data analysis. \emph{Microscopy and Microanalysis} \textbf{2021}, \emph{27}, 712--743\relax
\mciteBstWouldAddEndPuncttrue
\mciteSetBstMidEndSepPunct{\mcitedefaultmidpunct}
{\mcitedefaultendpunct}{\mcitedefaultseppunct}\relax
\EndOfBibitem
\bibitem[Jiang \latin{et~al.}(2022)Jiang, Parsonnet, Qualls, Zhao, Susarla, Pesquera, Dasgupta, Acharya, Zhang, Gosavi, Lin, Nikonov, Li, Young, Ramesh, and Martin]{jiang2022enabling}
Jiang,~Y. \latin{et~al.}  Enabling ultra-low-voltage switching in \ce{BaTiO3}. \emph{Nature Materials} \textbf{2022}, \emph{21}, 779--785\relax
\mciteBstWouldAddEndPuncttrue
\mciteSetBstMidEndSepPunct{\mcitedefaultmidpunct}
{\mcitedefaultendpunct}{\mcitedefaultseppunct}\relax
\EndOfBibitem
\bibitem[Joy \latin{et~al.}(1993)Joy, Zhang, Zhang, Hashimoto, Bunn, Allard, and Nolan]{joy1993practical}
Joy,~D.~C.; Zhang,~Y.-S.; Zhang,~X.; Hashimoto,~T.; Bunn,~R.; Allard,~L.; Nolan,~T. Practical aspects of electron holography. \emph{Ultramicroscopy} \textbf{1993}, \emph{51}, 1--14\relax
\mciteBstWouldAddEndPuncttrue
\mciteSetBstMidEndSepPunct{\mcitedefaultmidpunct}
{\mcitedefaultendpunct}{\mcitedefaultseppunct}\relax
\EndOfBibitem
\bibitem[Kirkland(2010)]{Kirkland_2010}
Kirkland,~E.~J. \emph{Advanced Computing in Electron Microscopy}; Springer {US}, 2010\relax
\mciteBstWouldAddEndPuncttrue
\mciteSetBstMidEndSepPunct{\mcitedefaultmidpunct}
{\mcitedefaultendpunct}{\mcitedefaultseppunct}\relax
\EndOfBibitem
\bibitem[Kruse \latin{et~al.}(2006)Kruse, Schowalter, Lamoen, Rosenauer, and Gerthsen]{kruse2006determination}
Kruse,~P.; Schowalter,~M.; Lamoen,~D.; Rosenauer,~A.; Gerthsen,~D. Determination of the mean inner potential in {III}--{V} semiconductors, {Si} and {Ge} by density functional theory and electron holography. \emph{Ultramicroscopy} \textbf{2006}, \emph{106}, 105--113\relax
\mciteBstWouldAddEndPuncttrue
\mciteSetBstMidEndSepPunct{\mcitedefaultmidpunct}
{\mcitedefaultendpunct}{\mcitedefaultseppunct}\relax
\EndOfBibitem
\bibitem[Han \latin{et~al.}(2014)Han, Marshall, Wu, Schofield, Aoki, Twesten, Hoffman, Walker, Ahn, and Zhu]{han2014interface}
Han,~M.-G.; Marshall,~M.~S.; Wu,~L.; Schofield,~M.~A.; Aoki,~T.; Twesten,~R.; Hoffman,~J.; Walker,~F.~J.; Ahn,~C.~H.; Zhu,~Y. Interface-induced nonswitchable domains in ferroelectric thin films. \emph{Nature Communications} \textbf{2014}, \emph{5}, 4693\relax
\mciteBstWouldAddEndPuncttrue
\mciteSetBstMidEndSepPunct{\mcitedefaultmidpunct}
{\mcitedefaultendpunct}{\mcitedefaultseppunct}\relax
\EndOfBibitem
\bibitem[Toyama \latin{et~al.}(2020)Toyama, Seki, Anada, Sasaki, Yamamoto, Ikuhara, and Shibata]{toyama2020quantitative}
Toyama,~S.; Seki,~T.; Anada,~S.; Sasaki,~H.; Yamamoto,~K.; Ikuhara,~Y.; Shibata,~N. Quantitative electric field mapping of a p--n junction by {DPC STEM}. \emph{Ultramicroscopy} \textbf{2020}, \emph{216}, 113033\relax
\mciteBstWouldAddEndPuncttrue
\mciteSetBstMidEndSepPunct{\mcitedefaultmidpunct}
{\mcitedefaultendpunct}{\mcitedefaultseppunct}\relax
\EndOfBibitem
\bibitem[da~Silva \latin{et~al.}(2022)da~Silva, Sadre~Momtaz, Monroy, Okuno, Rouviere, Cooper, and Den~Hertog]{da2022assessment}
da~Silva,~B.~C.; Sadre~Momtaz,~Z.; Monroy,~E.; Okuno,~H.; Rouviere,~J.-L.; Cooper,~D.; Den~Hertog,~M.~I. Assessment of Active Dopants and p--n Junction Abruptness Using In Situ Biased {4D-STEM}. \emph{Nano Letters} \textbf{2022}, \emph{22}, 9544--9550\relax
\mciteBstWouldAddEndPuncttrue
\mciteSetBstMidEndSepPunct{\mcitedefaultmidpunct}
{\mcitedefaultendpunct}{\mcitedefaultseppunct}\relax
\EndOfBibitem
\bibitem[Pintilie and Alexe(2005)Pintilie, and Alexe]{pintilie2005metal}
Pintilie,~L.; Alexe,~M. Metal-ferroelectric-metal heterostructures with Schottky contacts. I. Influence of the ferroelectric properties. \emph{Journal of Applied Physics} \textbf{2005}, \emph{98}\relax
\mciteBstWouldAddEndPuncttrue
\mciteSetBstMidEndSepPunct{\mcitedefaultmidpunct}
{\mcitedefaultendpunct}{\mcitedefaultseppunct}\relax
\EndOfBibitem
\bibitem[Nguyen \latin{et~al.}(2005)Nguyen, Davydov, Chandler-Horowitz, and Frank]{nguyen2005sub}
Nguyen,~N.~V.; Davydov,~A.~V.; Chandler-Horowitz,~D.; Frank,~M.~M. Sub-bandgap defect states in polycrystalline hafnium oxide and their suppression by admixture of silicon. \emph{Applied Physics Letters} \textbf{2005}, \emph{87}\relax
\mciteBstWouldAddEndPuncttrue
\mciteSetBstMidEndSepPunct{\mcitedefaultmidpunct}
{\mcitedefaultendpunct}{\mcitedefaultseppunct}\relax
\EndOfBibitem
\bibitem[Ramo \latin{et~al.}(2007)Ramo, Gavartin, Shluger, and Bersuker]{ramo2007spectroscopic}
Ramo,~D.~M.; Gavartin,~J.; Shluger,~A.; Bersuker,~G. Spectroscopic properties of oxygen vacancies in monoclinic \ce{HfO2} calculated with periodic and embedded cluster density functional theory. \emph{Physical Review B} \textbf{2007}, \emph{75}, 205336\relax
\mciteBstWouldAddEndPuncttrue
\mciteSetBstMidEndSepPunct{\mcitedefaultmidpunct}
{\mcitedefaultendpunct}{\mcitedefaultseppunct}\relax
\EndOfBibitem
\bibitem[Islamov \latin{et~al.}(2019)Islamov, Gritsenko, Perevalov, Pustovarov, Orlov, Chernikova, Markeev, Slesazeck, Schroeder, Mikolajick, and Krasnikov]{islamov2019identification}
Islamov,~D.~R.; Gritsenko,~V.~A.; Perevalov,~T.~V.; Pustovarov,~V.~A.; Orlov,~O.~M.; Chernikova,~A.~G.; Markeev,~A.~M.; Slesazeck,~S.; Schroeder,~U.; Mikolajick,~T.; Krasnikov,~G.~Y. Identification of the nature of traps involved in the field cycling of \ce{Hf_{0.5}Zr_{0.5}O_{2}}-based ferroelectric thin films. \emph{Acta Materialia} \textbf{2019}, \emph{166}, 47--55\relax
\mciteBstWouldAddEndPuncttrue
\mciteSetBstMidEndSepPunct{\mcitedefaultmidpunct}
{\mcitedefaultendpunct}{\mcitedefaultseppunct}\relax
\EndOfBibitem
\bibitem[Kresse and Furthm{\"u}ller(1996)Kresse, and Furthm{\"u}ller]{kresse1996efficiency}
Kresse,~G.; Furthm{\"u}ller,~J. Efficiency of ab-initio total energy calculations for metals and semiconductors using a plane-wave basis set. \emph{Computational Materials Science} \textbf{1996}, \emph{6}, 15--50\relax
\mciteBstWouldAddEndPuncttrue
\mciteSetBstMidEndSepPunct{\mcitedefaultmidpunct}
{\mcitedefaultendpunct}{\mcitedefaultseppunct}\relax
\EndOfBibitem
\bibitem[Kresse and Furthm{\"u}ller(1996)Kresse, and Furthm{\"u}ller]{kresse1996efficient}
Kresse,~G.; Furthm{\"u}ller,~J. Efficient iterative schemes for ab initio total-energy calculations using a plane-wave basis set. \emph{Physical review B} \textbf{1996}, \emph{54}, 11169\relax
\mciteBstWouldAddEndPuncttrue
\mciteSetBstMidEndSepPunct{\mcitedefaultmidpunct}
{\mcitedefaultendpunct}{\mcitedefaultseppunct}\relax
\EndOfBibitem
\bibitem[Bl{\"o}chl(1994)]{blochl1994projector}
Bl{\"o}chl,~P.~E. Projector augmented-wave method. \emph{Physical Review B} \textbf{1994}, \emph{50}, 17953\relax
\mciteBstWouldAddEndPuncttrue
\mciteSetBstMidEndSepPunct{\mcitedefaultmidpunct}
{\mcitedefaultendpunct}{\mcitedefaultseppunct}\relax
\EndOfBibitem
\bibitem[Kresse and Joubert(1999)Kresse, and Joubert]{kresse1999ultrasoft}
Kresse,~G.; Joubert,~D. From ultrasoft pseudopotentials to the projector augmented-wave method. \emph{Physical Review B} \textbf{1999}, \emph{59}, 1758\relax
\mciteBstWouldAddEndPuncttrue
\mciteSetBstMidEndSepPunct{\mcitedefaultmidpunct}
{\mcitedefaultendpunct}{\mcitedefaultseppunct}\relax
\EndOfBibitem
\bibitem[Perdew \latin{et~al.}(1996)Perdew, Burke, and Ernzerhof]{perdew1996generalized}
Perdew,~J.~P.; Burke,~K.; Ernzerhof,~M. Generalized gradient approximation made simple. \emph{Physical Review Letters} \textbf{1996}, \emph{77}, 3865\relax
\mciteBstWouldAddEndPuncttrue
\mciteSetBstMidEndSepPunct{\mcitedefaultmidpunct}
{\mcitedefaultendpunct}{\mcitedefaultseppunct}\relax
\EndOfBibitem
\bibitem[Monkhorst and Pack(1976)Monkhorst, and Pack]{monkhorst1976special}
Monkhorst,~H.~J.; Pack,~J.~D. Special points for {Brillouin-zone} integrations. \emph{Physical review B} \textbf{1976}, \emph{13}, 5188\relax
\mciteBstWouldAddEndPuncttrue
\mciteSetBstMidEndSepPunct{\mcitedefaultmidpunct}
{\mcitedefaultendpunct}{\mcitedefaultseppunct}\relax
\EndOfBibitem
\bibitem[Wang \latin{et~al.}(2021)Wang, Xu, Liu, Tang, and Geng]{wang2021vaspkit}
Wang,~V.; Xu,~N.; Liu,~J.-C.; Tang,~G.; Geng,~W.-T. {VASPKIT: A} user-friendly interface facilitating high-throughput computing and analysis using {VASP} code. \emph{Computer Physics Communications} \textbf{2021}, \emph{267}, 108033\relax
\mciteBstWouldAddEndPuncttrue
\mciteSetBstMidEndSepPunct{\mcitedefaultmidpunct}
{\mcitedefaultendpunct}{\mcitedefaultseppunct}\relax
\EndOfBibitem
\bibitem[10.5281/zenodo.6659919()]{hyperspy}
Hyperspy\relax
\mciteBstWouldAddEndPuncttrue
\mciteSetBstMidEndSepPunct{\mcitedefaultmidpunct}
{\mcitedefaultendpunct}{\mcitedefaultseppunct}\relax
\EndOfBibitem
\end{mcitethebibliography}

\end{document}